\documentclass[prb,notitlepage]{revtex4-1}

\usepackage{amsmath}   
\usepackage{graphicx}   
\usepackage{color}
\usepackage{verbatim}
\usepackage{amssymb}
\usepackage{amsmath}
\usepackage{epsfig}

\usepackage[dvipsnames]{xcolor}

\usepackage{cancel}
\usepackage{verbatim}
\usepackage{mathtools}

\def\bb#1{\textcolor{blue}{#1}}

\begin{document}

\title{Spin-orbit gravitational locking -- an effective potential approach}

\author{
Christopher Clouse$^{1}$, Andrea Ferroglia$^{2,3}$ and Miguel C. N. Fiolhais$^{1,3,4}$
\\[3mm]
{\footnotesize {\it 
$^1$ Science Department, Borough of Manhattan Community College, The City University of New York, \\ 
     199 Chambers St, New York, NY 10007, USA \\
$^2$ Physics Department, New York City College of Technology, The City University of New York, 300 Jay Street, Brooklyn, NY 11201, USA \\
$^3$ The Graduate School and University Center, The City University of New York, 365 Fifth Avenue, New York, NY 10016, USA \\
$^4$ LIP, Physics Department, University of Coimbra, 3004-516 Coimbra, Portugal\\
}}
}

\begin{abstract}

The objective of this paper is to study the  tidally locked 3:2 spin–orbit resonance of Mercury around the Sun. In order to achieve this goal, the effective potential energy that determines the spinning motion of an ellipsoidal planet around its axis is considered. By studying the rotational potential energy of an ellipsoidal planet orbiting a spherical star on an elliptic orbit with fixed eccentricity and semi-major axis, it is shown that the system presents an infinite number of metastable equilibrium configurations. These states correspond to local minima of the rotational potential energy averaged over an orbit, where the ratio between the rotational period of the planet around its axis and the revolution period around the star is fixed. The configurations in which this ratio is an integer or an half integer are of particular interest. Among these configurations, the deepest minimum in the average potential energy corresponds  to a situation where the rotational and orbital motion of the planet are synchronous, and the system is tidally locked. The next-to-the deepest minimum corresponds to the case in which the planet rotates three times around its axis in the time that it needs to complete two orbits around the Sun. The latter is indeed the case that describes Mercury's motion. The method discussed in this work allows one to identify the integer and half-integer ratios that correspond to spin-orbit resonances and to describe the motion of the planet in the resonant orbit. \\

\noindent This Accepted Manuscript is available for reuse under a CC BY-NC-ND license after the 12 month embargo period provided that all the terms and conditions of the license are adhered to.

\end{abstract}

\maketitle

\section{Introduction}


Tidal locking between two astronomical objects is a well known phenomenon that has fascinated physicists, philosophers and astronomers throughout centuries~\cite{arons,gron,white,withers,koenders,butikov,razmi,massi,urbassek,pujol,ng,cregg,norsen}. In a tidally locked two-body system, the orbital angular velocity of the objects around the common center of mass of {the} system is equal to the angular speed of one or both objects spinning around their own axes. The most noticeable example is the case of the Moon orbiting around planet Earth. In this case, the Moon rotates  in an approximately circular orbit around the center-of-mass of the Earth-Moon system in exactly the same time as it takes to {revolve around its axis}. Consequently, the near side of the Moon is always facing the Earth, while the far side is always hidden from an Earthling's view. If the Earth and the Moon are considered as an isolated dynamical system, the tidal friction resulting from the bulges produced by gravitational force of the Moon on Earth’s crust would eventually dissipate energy and slow down Earth’s rotation until the system is completely tidally locked, \emph{i.e.} both the Moon and the Earth would spin around their axes in the same time that it would take them to orbit around the center of mass of the system.

While the conceptual and mathematical understanding of tidal interactions between astronomical objects dates back to Johannes Kepler~\cite{kepler}, Sir Isaac Newton~\cite{newton} and Immanuel Kant~\cite{kant}, it has been shown during the past fifty years that the effect of tidal locking can be mathematically derived using an effective potential approach~\cite{kopal72,counselman73,vanhamme79,hut80,mcdonald,ferroglia}. In this framework, tidal locking is obtained by minimizing the effective potential energy of the astronomical objects orbiting each other, taking into account their rotational kinetic energies around their own axes, and assuming the total angular momentum of the system remains constant. The local minimum of the effective potential of the two-body system corresponds to a  circular orbit  configuration in which the two objects are completely tidally locked to each other. Moreover, this approach is also useful to study the stability of the system. The existence of a local minimum in the effective potential, \emph{i.e.} a stable tidally-locked configuration, depends on a single dimensionless parameter, corresponding to the case of a fold catastrophe in catastrophe theory\cite{guemez,fiolhais1,fiolhais2}. This control parameter is {a function of} the objects’ masses and moments of inertia and the total angular momentum of the system.

Despite the fact that tidal locking had already been studied in some detail, in 1964 the recently decommissioned Arecibo Telescope revealed a surprising new manifestation of a closely related phenomenon. The rotation period of Mercury around its own axis {is} only 59 days, as opposed to its 88-day orbital period~\cite{dyce} - an approximate 3:2 spin-orbit resonance. The reason for this anomalous behavior was soon identified as stemming from the ellipsoid shape of Mercury and its high eccentricity orbit around the Sun. 
This spin-orbit resonance due to Mercury's ellipsoidal shape is stabilized by the tidal torque applied by the Sun on Mercury.
In this paper, this result is obtained in a pedagogical manner by considering the potential energy that regulates the rotational motion of the planet around its axis, where the metastable equilibrium configurations appear as local minima if the ellipsoidal satellite orbits around a central spherical object in an elliptic orbit.

In order to derive spin-orbit resonances in the aforementioned case, the total energy of the system is calculated in Section~\ref{sec:energy}, taking into account the quadrupole correction to the gravitational potential energy of an ellipsoidal shaped planet orbiting around a spherical central star. In Section~{\ref{sec:eqforgamma}} it is shown how for integer and half-integer values of the ratio between the rotational and orbital periods of the planet, {a certain angle $\gamma$, that depends on the planet's rotational angle and on the mean anomaly of the orbital motion}, satisfies an equation of the same type as the equation of motion for a simple pendulum. It is then shown that in these conditions the rotational and orbital periods of the planet can remain in a fixed (half-)integer ratio.  
In Section~\ref{sec:rotpoten} the situation is reanalyzed by considering the shape of the rotational potential energy averaged over the orbital period. This analysis shows that for integer and half-integer ratios of the orbital period, the averaged rotational potential energy shows metastable minima for $\gamma = 0$ or $\gamma = \pi/2$.
Conclusions are drawn in Section~\ref{sec:conclusions}.

\section{Energy of the system}
\label{sec:energy}

\begin{figure}
    \centering
\includegraphics[width=0.70\textwidth]{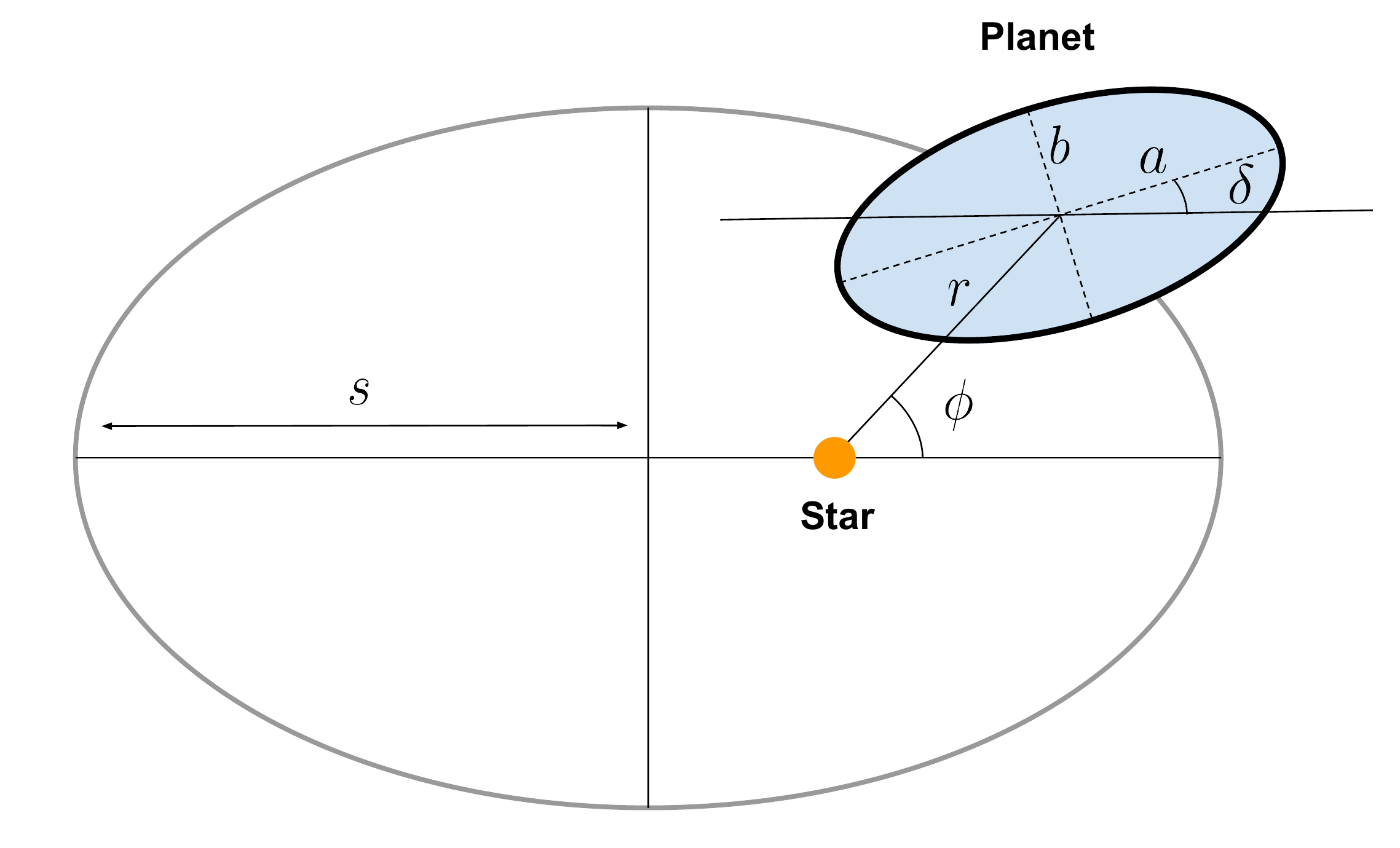}
    \caption{Schematic representation (not to scale) of the planet orbiting the star along an elliptic orbit. The planet also spins {around an axis  perpendicular to the plane of the orbit and going through its center}. The figure shows the relation between the angle $\phi$ and the angle $\delta$. }
    \label{fig:Spin-Orbit}
\end{figure}

Consider a planet of mass $m$ orbiting a star of mass $M_s$. The star is modeled  as a perfectly spherical object, while the planet is an ellipsoid of semi-axes of length $a > b > c$. 
The rotational velocity of the star around its axis does not play a role in the argument exposed below, so it is taken equal to zero for simplicity. On the contrary, the planet is assumed to be spinning about an axis perpendicular to the orbital plane. This axis {of rotation} is supposed to coincide with {the shortest axis} of symmetry of the ellipsoid. The gravitational potential energy of this system can be studied through a multipole expansion, {as discussed for example in Sussman and Wisdom's book~\cite{sussman}}. By truncating the expansion after the quadrupole contribution, the gravitational potential can be written as
\begin{equation} \label{eq:grpotential}
    U_{\text{gr}} = - \frac{G M_s m}{r} - \frac{3 G M_s }{2 r^3}  \left[ \left(B -A\right) \cos^2 \left(  \phi - \delta \right) - \frac{1}{3} \left(2 B -A - C \right)  \right] + {\mathcal{O}} \left(\frac{1}{r^4} \right)\, ,
\end{equation}
where $r$ is the distance between the star and the center of the planet. The angle $\phi$ is the angle that the line joining the star to the planet makes with the $x$-axis of the frame of reference, which is taken as centered on the star, since the star is assumed to be much more massive than the planet, $m \ll M_s$.
The semi-axes $a$ and $b$ of the planet lie on the orbital plane, which is assumed to be the $x-y$ plane. The angle that the longest axis $a$ makes with the horizontal direction is indicated with $\delta$ in Eq.~(\ref{eq:grpotential}). 
In the case of an elliptic orbit, {in which the $x$-axis can be conveniently chosen along the line joining the aphelion and the perihelion}, the situation is sketched in Figure~\ref{fig:Spin-Orbit}.
The quantities $A,B$ and $C$ are the moments of inertia of the ellipsoid with respect to its principal axes
\begin{equation}
    A = \frac{m}{5} \left(b^2+c^2 \right) \, , \qquad 
    B = \frac{m}{5} \left(a^2+c^2 \right) \, , \qquad
    C = \frac{m}{5} \left(a^2+b^2 \right)  \, .
\end{equation}
 {Indeed, Eq.~(\ref{eq:grpotential}) coincides with Eq.~(2.74) in~\cite{sussman}, if one considers that in the latter equation one should set $\alpha = \cos(\phi-\delta)$, $\beta =\sin(\phi-\delta)$, and $\gamma = 0$ to describe the configuration considered in Figure~\ref{fig:Spin-Orbit}.} 
 
For the purposes of this work it is possible to set $c = b$, which implies $C = B$. With this assumption $2 B- A-C  = B-A $ so that the gravitational potential in Eq.~(\ref{eq:grpotential}) has the simpler form
\begin{equation} \label{eq:grpotentialBC}
    U_{\text{gr}} = - \frac{G M_s m}{r} -\frac{3  G M_s }{2 r^3}  \left(B -A\right)   \left[ \cos^2 \left(  \phi - \delta \right) - \frac{1}{3}   \right] + {\mathcal{O}} \left(\frac{1}{r^4} \right)\, .
\end{equation}

The total mechanical energy of a system of two masses $M_s$ and $m$ orbiting each other under the potential in Eq.~(\ref{eq:grpotentialBC}) {is the sum of the orbital kinetic energy, the rotational kinetic energy due to the spinning of the ellipsoid, and of the gravitational potential energy;} it can be expressed {(in the c.o.m. frame)} as
\begin{equation} \label{eq:4}
    E = \frac{1}{2} m \left( \dot{r}^2 +r^2 \dot{\phi}^2 \right)  + \frac{1}{2} B \dot{\delta}^2  - \frac{G M_s m}{r} -\frac{3  G M_s }{2 r^3}  \left(B -A\right)   \left[ \cos^2 \left(  \phi - \delta \right) - \frac{1}{3}   \right] \, .
\end{equation} 
%
{The planet's angular velocity around its axis is $\dot{\delta}$.} The angular velocity of the planet in its orbit around the star is indicated by $\dot{\phi}$. {The angular momentum of the system at a given instant in time}  is
\begin{equation} \label{eq:angmom}
L \equiv  m r^2 \dot{\phi}  + B \dot{\delta} \, ,
\end{equation}
{where the first term in the r.h.s. of Eq.~(\ref{eq:angmom}) is the orbital angular momentum, while the second term is the angular momentum associated to the rotation of the ellipsoidal planet around an axis of length $2 b$. If the planet follows an elliptic orbit, the orbital angular momentum is conserved, so that it is convenient to set}
\begin{equation} \label{eq:l}
    l \equiv m r^2 \dot{\phi} \, .
\end{equation}
Consequently, {by solving Eq.~(\ref{eq:l}) w.r.t. $\dot{\phi}$ and by inserting the result in Eq.~(\ref{eq:4})},  the  total mechanical energy of the system can then be expressed as
\begin{equation} \label{eq:energyrpd}
    E = \frac{1}{2} m \dot{r}^2 - G \frac{M_s m}{r} 
    -\frac{3  G M_s }{2 r^3}  \left(B -A\right)   \left[ \cos^2 \left(  \phi - \delta \right) - \frac{1}{3}   \right]
    +\frac{l^2}{2 m r^2} + {\frac{1}{2} B \dot{\delta}^2} \, .
\end{equation}

\boldmath
\section{Spin-orbit gravitational locking} \label{sec:eqforgamma}
\unboldmath

\begin{figure}
    \centering
\includegraphics[width=0.55\textwidth]{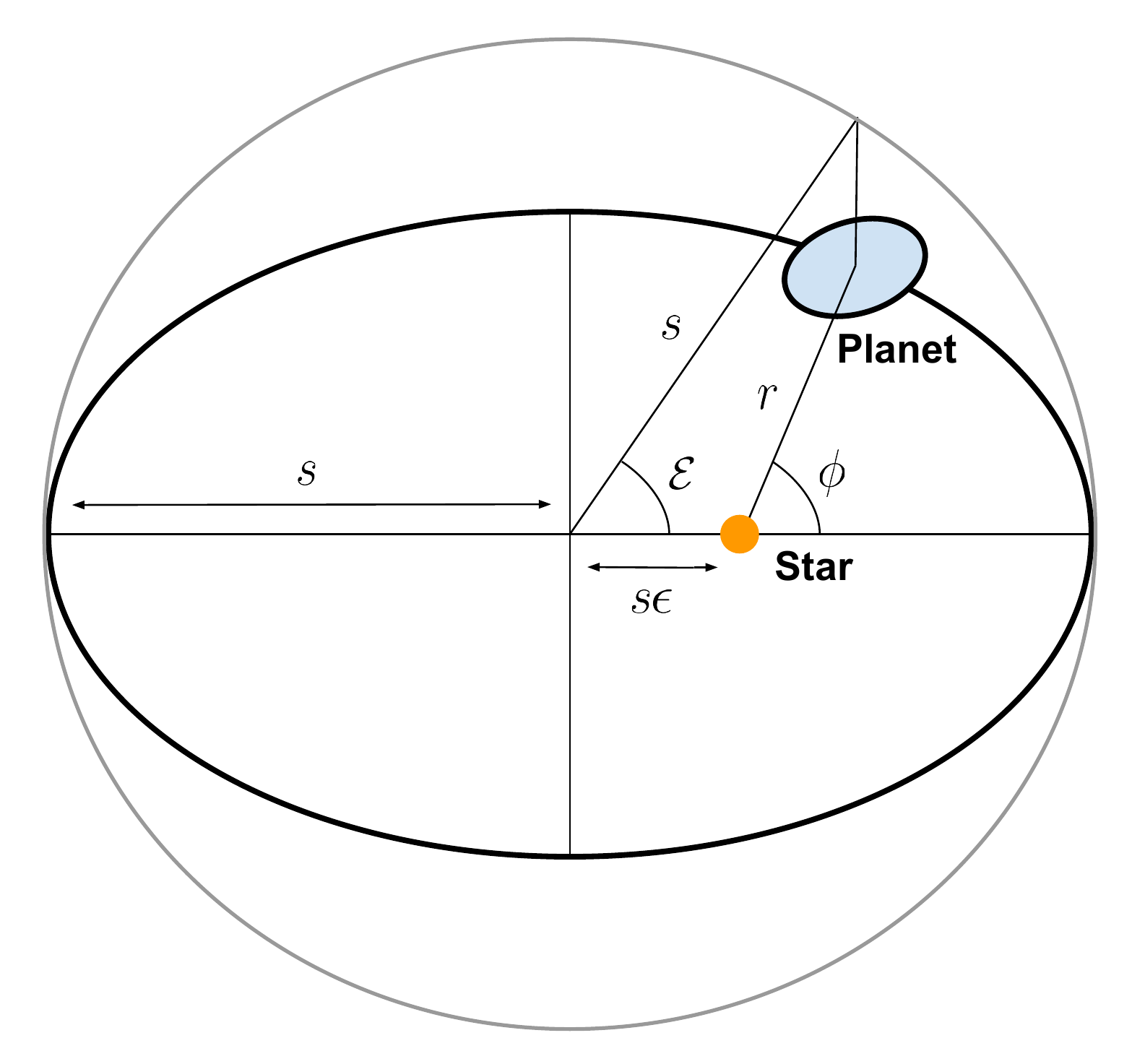}
    \caption{Relation between the elliptic anomaly ${\mathcal E}$ and the {true anomaly} $\phi$. }
    \label{fig:elliptic-anomaly}
\end{figure}

The orbits of the planets of the solar system are described to an excellent accuracy by ellipses. For elliptic orbits,  there is a fixed relation between the distance between the Sun and the planet, $r$,  and the {true anomaly} $\phi$:

\begin{equation} \label{eq:ellipse}
    r = \frac{s \left(1- \epsilon^2 \right)}{1 + \epsilon \cos \phi} \, ,
\end{equation}
where $\epsilon$ is the orbit eccentricity and $s$ is the length of the semi-major axis of the orbit.
In addition, the part of the mechanical energy of the system associated to the orbital motion is fixed and it depends on the eccentricity and  semi-major axis of the orbit: 
\begin{equation}\label{eq:orbE}
E_{\text{orb}} = \frac{1}{2} m \dot{r}^2 - G \frac{M_s m}{r} + \frac{l^2}{2 m r^2}  = \frac{G^2 M_s^2 m^3}{2 l^2} \left(\epsilon^2-1 \right)\, ,
\end{equation}
where, since $0< \epsilon < 1$ in an elliptic orbit, $E_{\text{orb}} < 0$. {The last equality in Eq.~(\ref{eq:orbE}) can be obtained by observing that at the planet perihelion, $\phi = 0$, so that $r = s (1-\epsilon)$; in addition at the perihelion the orbital energy of the planet coincides with its effective orbital potential energy, since its radial velocity is zero. Therefore, the last equality in Eq.~(\ref{eq:orbE}) can be verified by calculating the effective potential energy at the perihelion and by using the relation between angular momentum, semi-major axis and eccentricity of the orbit~\cite{taylor}.} {An expression for the orbital energy as a function of the semi-major axis rather than eccentricity can be found in~\cite{barger}; such expression is equivalent to the r.h.s. of Eq.~(\ref{eq:orbE}).} By using Eq.~(\ref{eq:energyrpd}), one finds 
\begin{equation} \label{eq:rotE}
E_{\text{rot}} = E - E_{\text{orb}} = 
\frac{1}{2} B \dot{\delta}^2 -\frac{3  G M_s }{2 r^3}  \left(B -A\right)   \left[ \cos^2 \left(  \phi - \delta \right) - \frac{1}{3}   \right] \, ,
\end{equation}
where, since the orbit is elliptic, the rotational energy $E_{\text{rot}}$ can be rewritten {by using Eq.~(\ref{eq:ellipse}) to replace $r$ in Eq.~(\ref{eq:rotE})} 
\begin{equation}
    E_{\text{rot}} =\frac{1}{2} B \dot{\delta}^2 - \frac{3  G M_s }{2 s^3}  \left(B -A\right)  \left(\frac{1 +\epsilon \cos \phi}{1 - \epsilon^2} \right)^3 \left[ \cos^2 \left(  \phi - \delta \right) - \frac{1}{3}   \right] \, .
\end{equation}
In addition, to simplify several of the equations that  appear later on in this work, it is useful to introduce the quantity
\begin{equation}
    Q \equiv \frac{3  G M_s }{2 s^3}  \left(B -A\right) \, .
\end{equation}
Furthermore, following \cite{goldreich_peale}, it is now convenient to introduce the angle 
\begin{equation} \label{eq:gamma}
    \gamma  \equiv \delta - p M \, ,
\end{equation}
where $M$ is the mean anomaly of the orbital motion~\cite{,Montenbruck,Meeus} and $p$ is a generic dimensionless parameter. By taking the time derivative of Eq.~(\ref{eq:gamma}) one finds $\dot{\gamma} = \dot{\delta} - p n$, where $n = 2 \pi/ T$ is the mean motion and $T$ the orbital period. If $\dot{\delta}$ remains close to $p n$ through every orbit, $\gamma$ is almost constant. In other words, in the case {where} the planet is spinning with a constant angular speed {that matches the product $p n$, then} there is a constant phase shift between the angles $\delta$ and $pM$. {This behavior can be studied in detail by replacing} 
\begin{equation}
    \delta  = \gamma  + p M \, ,
\end{equation}
in Eq.~(\ref{eq:rotE}), and further replacing $\phi$ with its expression in terms of $\epsilon$ and $M$. 

In order to find the relation between $\phi$, $\epsilon$, and $M$, it is useful to start by observing that simple geometric considerations (see Figure~\ref{fig:elliptic-anomaly}) allow one to write a relation between the {true anomaly} $\phi$, the eccentricity $\epsilon$ and the elliptic anomaly 
${\mathcal E}$:
\begin{equation} \label{eq:phiintermsofE}
    \cos \phi = \frac{\cos {\mathcal E} -\epsilon}{1 - \epsilon \cos {\mathcal E}} \, .
\end{equation}
The relation in Eq.~(\ref{eq:phiintermsofE}) can be solved w.r.t. the elliptic anomaly to find
\begin{equation}\label{eq:Eintermsofphi}
    {\mathcal E} = \arccos{\left(\frac{\cos \phi + \epsilon}{1+  \epsilon \cos \phi} \right)} \, .
\end{equation}
Subsequently, it is possible to relate the elliptic anomaly to the mean anomaly through \emph{Kepler's equation}~\cite{danby}
{\begin{equation}
    \label{eq:kepler}
    M = {\mathcal E} - \epsilon \sin {\mathcal E} \, .
\end{equation}}
By combining Eq.~(\ref{eq:kepler}) with Eq.~(\ref{eq:Eintermsofphi}) it is possible to write $M$ as a function of $\phi$. {This function can be written as} a power series in the eccentricity: 
\begin{align}
    M &= \phi  -  2 \epsilon  \sin \phi + \left(\frac{3}{4}\epsilon^2 + \frac{1}{8} \epsilon^4 + \frac{3}{64} \epsilon^6 \right) \sin \left( 2\phi \right) - \left(\frac{1}{3}\epsilon^3 + \frac{1}{8} \epsilon^5 \right) \sin \left( 3 \phi \right) 
   \nonumber  \\
    & +\left(\frac{5}{32}\epsilon^4 + \frac{3}{32} \epsilon^6 \right) \sin \left( 4 \phi \right) - \frac{3}{40}\epsilon^5 \sin\left(5 \phi \right) +  \frac{7}{192}\epsilon^6 \sin\left(6 \phi \right) 
    +{\mathcal O}\left(\epsilon^7 \right) \, .
\end{align}
Consequently, the inverse relation {between the true anomaly and mean anomaly} can be written as a power series in $\epsilon$ as well:
\begin{align} \label{eq:phivsM}
    \phi &= M + \left( 2 \epsilon  - \frac{1}{4} \epsilon^3  + \frac{5}{96} \epsilon^5 \right) \sin{M} 
    +\left(\frac{5}{4} \epsilon^2 - \frac{11}{24} \epsilon^4 + \frac{17}{192} \epsilon^6\right) \sin{\left(2M\right)} + \left( \frac{13}{12} \epsilon^3  -\frac{43}{64} \epsilon^5 \right)\sin\left(3 M\right)
    \nonumber \\
    &+ \left(\frac{103}{96} \epsilon^4 - \frac{451}{480} \epsilon^6 \right)\sin\left(4 M\right)
    +\frac{1097}{960} \epsilon^5 \sin\left(5 M\right)+\frac{1223}{960} \epsilon^6 \sin\left(6 M\right)+ \mathcal{O} \left(\epsilon^7 \right)  \, .
\end{align}

{As mentioned previously, for the purposes of this work it is sufficient to consider the case in which $\dot{\gamma}$ is small.} {In this approximation,} after inserting Eq.~(\ref{eq:phivsM}) in Eq.~(\ref{eq:rotE}), it is then possible to integrate over the mean anomaly $M$ in order to average the rotational energy over a complete orbit while keeping $\gamma$ fixed.
{The integration with respect to the mean anomaly is equivalent to an integration with respect to time, since the mean anomaly is a linear function of time. In addition, averaging over the mean anomaly, an approach already followed by Goldreich and Peale~\cite{goldreich_peale},  is straightforward once the relation {between the true anomaly and mean anomaly}, Eq.~(\ref{eq:phivsM}), is known.}
One then finds 
\begin{equation} \label{eq:averageU}
     \frac{1}{2 \pi} \int_{0}^{2 \pi} d M E_{\text{rot}} = \frac{1}{2} B \left(\dot{\gamma} + p n\right)^2 + Q \left[S(\epsilon) + R(p,\epsilon) \cos\left( 2(p \pi +\gamma ) \right) \sin{\left(2 p \pi\right)} \right] \, .
\end{equation}
The functions $S$ and $R$ have the expansions
\begin{align} \label{eq:expSandR}
S(\epsilon) =& -\frac{1}{6} -\frac{1}{4} \epsilon^2 - \frac{5}{8} \epsilon^4  +\mathcal{O} \left( \epsilon^6 \right)\nonumber \\
R(p,\epsilon) =& \frac{1}{2 \pi} \Biggl[ \frac{1}{p-1} \left(-\frac{1}{2} +\frac{5}{4} \epsilon^2 - \frac{13}{32} \epsilon^4 \right) + \frac{1}{p - \frac{3}{2}} \left(- \frac{7}{4} \epsilon + \frac{123}{32} \epsilon^3 \right)
+\frac{1}{p -\frac{1}{2}} \left(\frac{1}{4} \epsilon - \frac{1}{32} \epsilon^3
\right)+ \frac{1}{p -\frac{5}{2}} \left(-\frac{845}{96} \epsilon^3 \right)
\nonumber \\
& 
+ \frac{1}{p +\frac{1}{2}} \left(-\frac{1}{96} \epsilon^3 \right)
+\frac{1}{p -3} \left(-\frac{533}{32} \epsilon^4 \right)
+\frac{1}{p -2} \left({-\frac{17}{4} \epsilon^2} +  \frac{115}{12} \epsilon^4 \right)
+\frac{1}{p +1} \left(-\frac{1}{48} \epsilon^4 \right)
\Biggr]+\mathcal{O} \left( \epsilon^5 \right)\,.
\end{align}
The first term in Eq.~(\ref{eq:averageU}) can be interpreted as {the} kinetic energy for the rotation of the planet around its axis, while the second term can be read as the potential energy for the same variable.

One can then build the Lagrangian for $\gamma$ 
\begin{equation} 
    \mathcal{L} = \frac{1}{2} B \left(\dot{\gamma} + pn \right)^2 - Q  \left[S(\epsilon) + R(p,\epsilon) \cos\left( 2(p \pi +\gamma ) \right) \sin{\left(2 p \pi\right)} \right] \, .
\end{equation}
The equation of motion for $\gamma$ is therefore
\begin{equation} \label{eq:eom}
    \frac{d}{dt} \left[B \left(\dot{\gamma} + p n\right) \right] - 2 Q   R(p,\epsilon)  \sin{\left(2 p \pi\right)} \sin\left( 2(p \pi +\gamma )\right) = 
    B \ddot{\gamma} - 2 Q   R(p,\epsilon)  \sin{\left(2 p \pi\right)} \sin\left( 2(p \pi +\gamma )\right) = 0 \,.
\end{equation}
{By using a standard trigonometric identity, Eq.~(\ref{eq:eom})} can be rewritten as 
\begin{equation} \label{eq:eom2}
B \ddot{\gamma} - 2 Q   R(p,\epsilon)  \sin{\left(2 p \pi\right)} \left[ \sin\left( 2p \pi\right)\cos( 2\gamma ) +  \cos\left( 2p \pi\right)\sin( 2\gamma )
\right]  = 0 \,.
\end{equation}
Since the function $R(p,\epsilon)$ has at most single poles for $p=k/2$ with $k \in {\mathbb{N}}$, {the product $R(p,\epsilon) \sin(2 p \pi)$ has a finite limit for $p \to k/2$. One can then observe that in the square bracket in Eq.~(\ref{eq:eom2}) the factor $\sin(2 p \pi)$ vanishes for $p = k/2$, while $\cos(2 p \pi) = \pm 1$ for $p =k/2$.}
Consequently, for integer and half-integer values of the parameter $p$ the term proportional to  $\cos (2 \gamma)$ in Eq.~(\ref{eq:eom2})  vanishes, while the term proportional to $\sin (2 \gamma)$ survives.
Therefore, for $p = k/2$ the equation of motion for $\gamma$ becomes 
\begin{equation} \label{eq:pendulumgamma}
B \ddot{\gamma} - 2 Q R \left(p,\epsilon \right) \sin \left( 2 p \pi \right) \cos \left( 2 p \pi \right)\sin \left(2 \gamma \right)  = 0 \, .
\end{equation}
Eq.~(\ref{eq:pendulumgamma}) is the same type of equation of motion that is satisfied by the angle between the vertical direction and the thread supporting a simple pendulum.
This is indeed the result outlined at the beginning of \cite{goldreich_peale}, and the function $H(p,\epsilon)$ defined in \cite{goldreich_peale} is related to the function $R(p,\epsilon)$ defined above through the equation
\begin{equation}
    H\left(p, \epsilon \right) = -2 R \left(p,\epsilon \right) \sin \left( 2 p \pi\right) \cos \left( 2 p \pi\right)\, .
\end{equation}
{The expansion of $H$ with  respect to $\epsilon$ can be easily obtained starting from the expansion in Eq.~(\ref{eq:expSandR}).}

Consequently, the equation of motion for $\gamma$ can be further rewritten as
\begin{equation} \label{eq:pendulumgamma2}
B \ddot{\gamma} + Q H \left(p,\epsilon \right) \sin \left(2 \gamma \right)  = 0 \, .
\end{equation}

{It is important to stress once more that Eq.~(\ref{eq:pendulumgamma2}) was obtained under the assumption that $\dot{\gamma}$ is small and that $p=k/2$.} If  $H$ in Eq.~(\ref{eq:pendulumgamma2}) is positive, the equation describes  the oscillations of $\gamma$ about $\gamma = 0$ {and}  the amplitude of the oscillation depends on the initial value of $\gamma$, {since Eq.~(\ref{eq:pendulumgamma2}) is of the same type as the pendulum equation for an arbitrary swinging angle}. {In addition,} if both the initial value of $\gamma$ and $\dot{\gamma}$ are small {and $H > 0$}, {one can replace $\sin(2 \gamma) \approx 2 \gamma$ in } Eq.~(\ref{eq:pendulumgamma2}) {that reduces to the equation of a motion of a harmonic oscillator.} {In that case, Eq.~(\ref{eq:pendulumgamma2})} implies that $\gamma$ {will remain} $\approx 0$ all the time and, consequently, ${\delta} \approx p M$  with $p = k/2$, so that the periods of the  spin and orbital motions are locked in an integer or half-integer ratio. Therefore, if the longest semiaxis of the planet (denoted by $a$) points toward the sun at perihelion, it is again pointing toward the sun after two orbital periods. In these two orbital periods the planet will have completed $2 p = k$ revolutions around its axis. The frequency of small oscillations {of $\gamma$ around the value $\gamma = 0$} is
\begin{equation}
    \omega  = \sqrt{\frac{2Q H\left(p,\epsilon\right)}{B}} \, .
\end{equation}
If $H < 0$ instead, it is possible to see that $\gamma$ oscillates around the value $\gamma = \pi/2$. In this case, if the shortest semiaxis of the planet (denoted by $b$) is pointing toward the sun at perihelion, this axis will return to point toward the sun after two orbital periods. Also in this case, in these two orbital periods, the planet will have completed    $2 p = k$ revolutions around its axis.

\section{Resonant orbits \label{sec:rotpoten}}

\begin{figure}
    \centering
\begin{tabular}{cc}
 \includegraphics[width=0.49\textwidth]{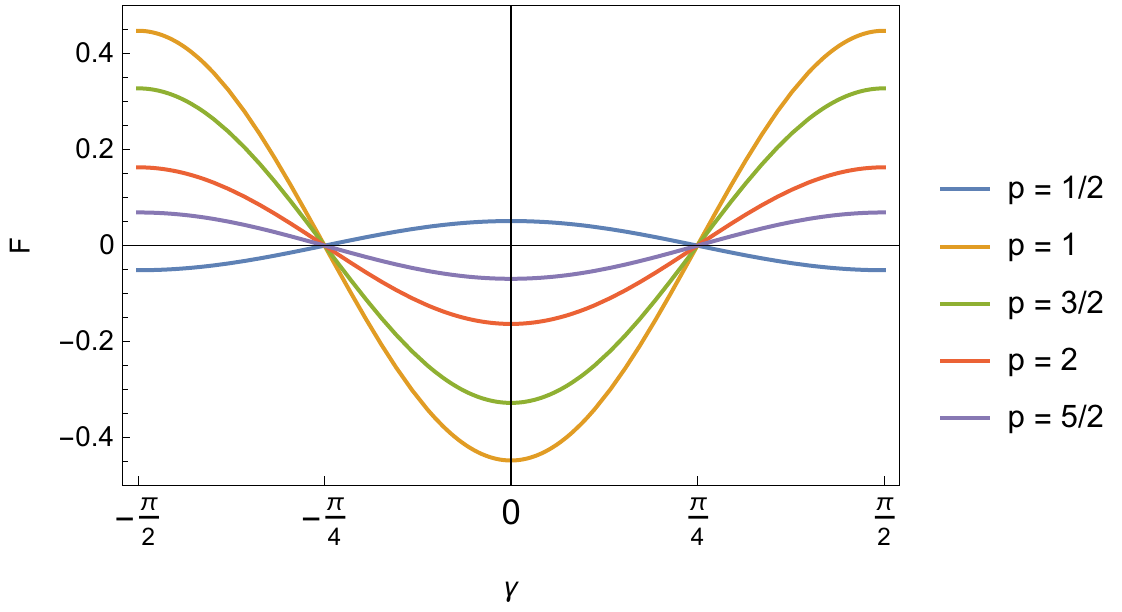}    & \includegraphics[width=0.49\textwidth]{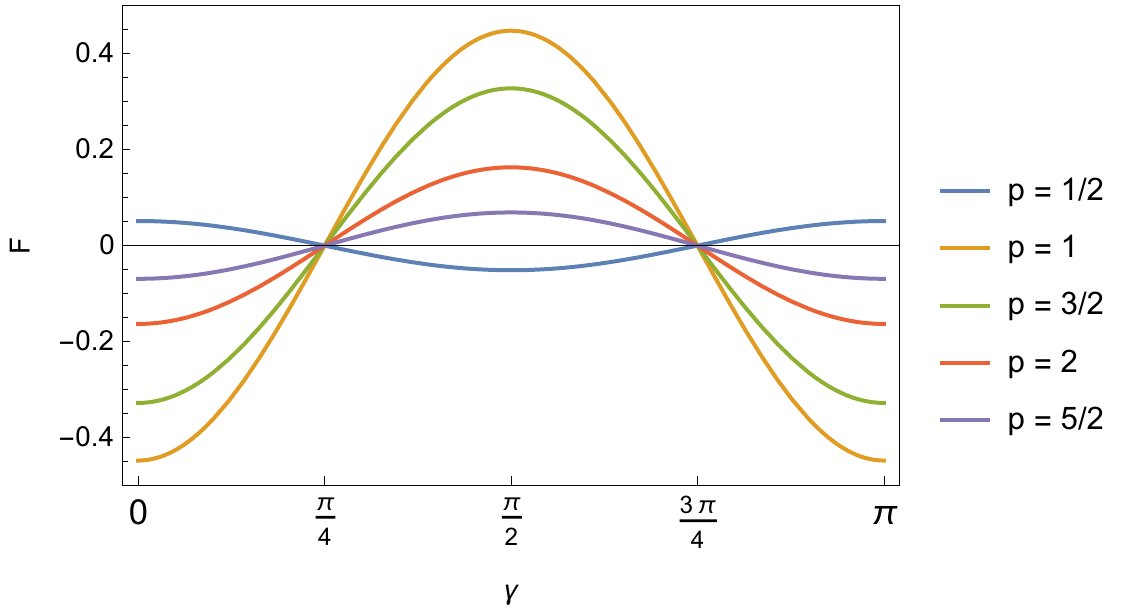}
\end{tabular}
    \caption{The function $F$ versus $\gamma$ for fixed values of $p = k/2$, for $\epsilon = 0.206$. }
    \label{fig:minimaofFgamma0}
\end{figure}

It is interesting to analyze the dependency of the potential energy term in Eq.~(\ref{eq:averageU}) with respect to the variable $\gamma$ and the parameter $p$. This dependence is encompassed in the function 
\begin{equation} \label{eq:F}
    F(p,\epsilon,\gamma) \equiv  R\left(p,\epsilon\right) \sin \left(2 p \pi \right) \cos\left(2 \left( p \pi + \gamma \right) \right) \, .
\end{equation}
By plotting the function $F$ with respect to $\gamma$, while keeping $p$ and $\epsilon$ fixed, it is possible to observe that for $p = k/2$ with $k \in {\mathbb N}$ (with the exception of $p=1/2$), there is a minimum of the function located at $\gamma = 0$. The location of the minimum of $F$ as a function of $\gamma$ is shown in the left panel of Figure~\ref{fig:minimaofFgamma0}. The depth of the minima of $F$ at $p = k/2$ depends on the order in $\epsilon$ at which the corresponding pole enters in the function $R\left(p,\epsilon \right)$, {as well as on the residue of the pole}.
{In particular, the function $F$ has a minimum at $\gamma = 0$ if the sign of the residue at a given $p = k/2$ in the function $R$ is negative.}
As expected, the deepest minimum corresponds to $p = 1$, since the function $R$ has a simple pole in $p = 1$ already at zeroth order in $\epsilon$ (see Eq.~(\ref{eq:expSandR})). The second deepest minimum {in the function $F$} is the one that corresponds to the Mercury spin-orbit resonance, $p = 3/2$. The simple pole at $p = 3/2$ appears at order $\epsilon$ in the expansion of the function $R$ in Eq.~(\ref{eq:expSandR}). {As the function $R$ has a simple pole at $p = 2$ whose residue is proportional to $\epsilon^3$, the third deepest minimum appears for $p=2$. The poles of other integer and half-integer values of $p$ in the function $R(p,\epsilon)$, that are proportional to higher orders of $\epsilon$, lead to shallower minima in $F(p,
\epsilon,\gamma)$ at $\gamma = 0$.}
%
\begin{figure}
    \centering
\begin{tabular}{cc}
 \includegraphics[width=0.50\textwidth]{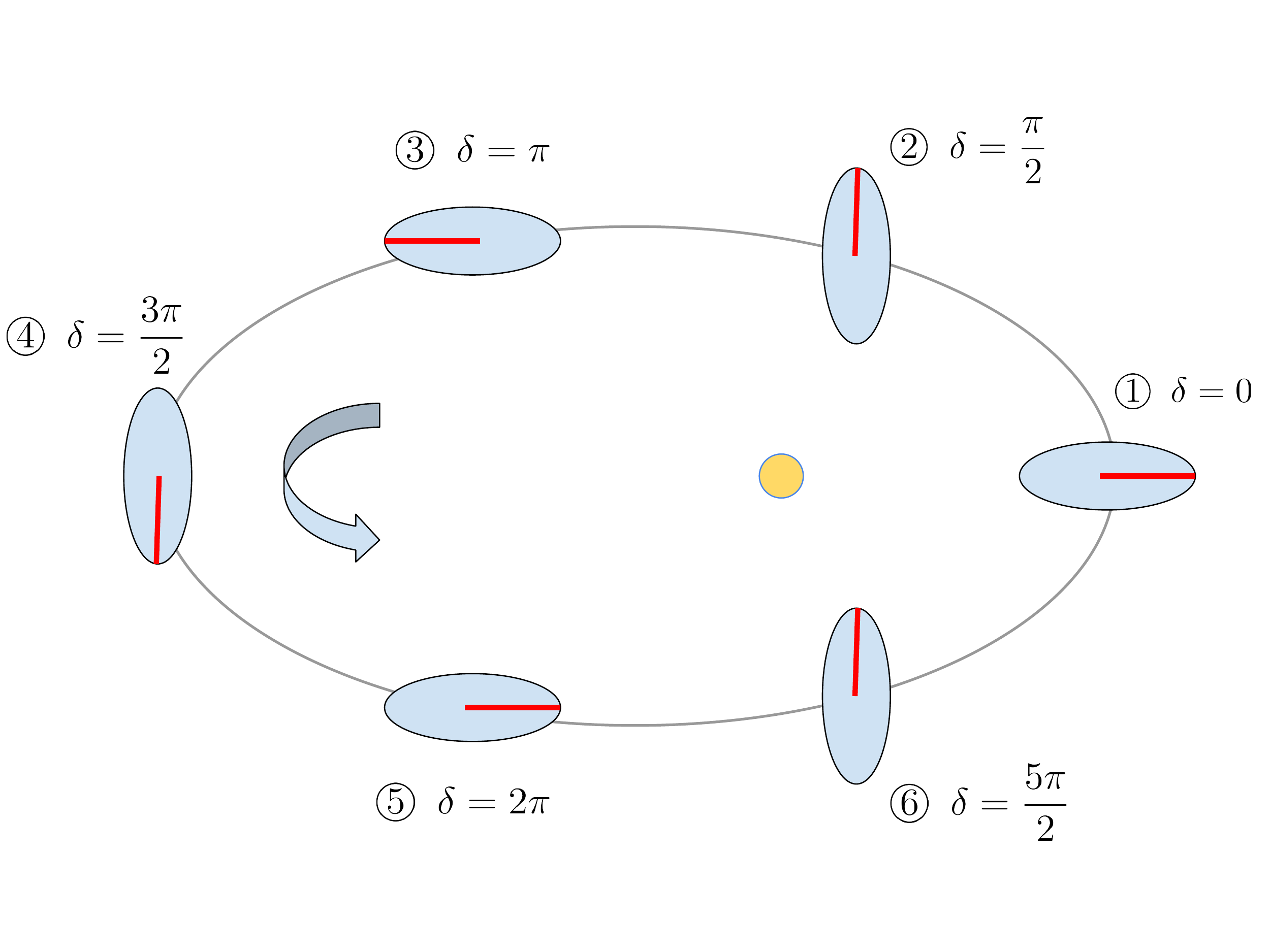}    & \includegraphics[width=0.50\textwidth]{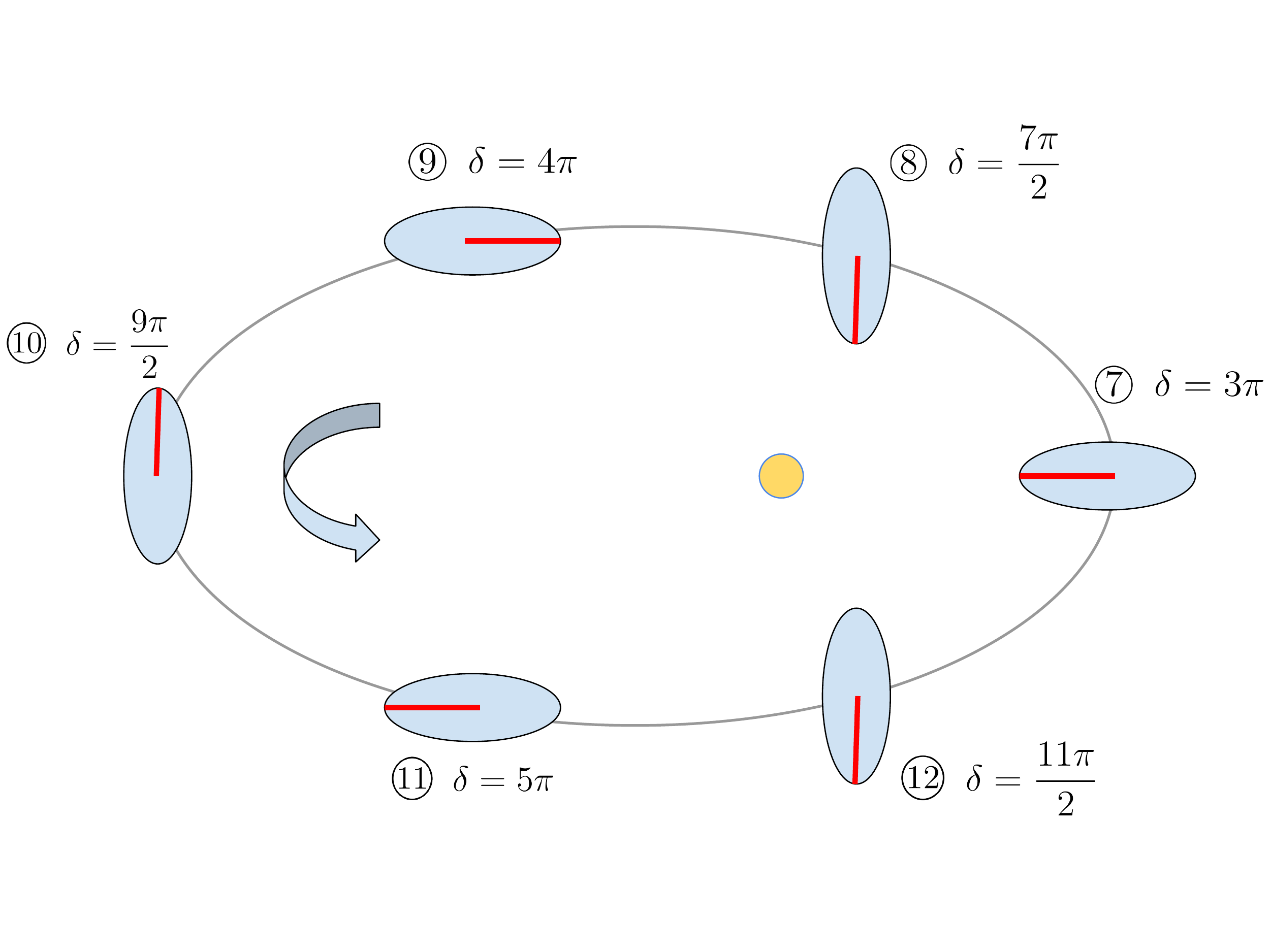}
\end{tabular}
    \caption{Schematic representation of the orbit of Mercury in a 3:2 spin-orbit resonance. The left panel refers to the the first orbit of the planet around the star, the right panel shows the second orbit of the planet around the star. A semi-major axis of the planet is drawn in red in order to show the angle of rotation of of the planet around its axis.}
    \label{fig:so3to2}
\end{figure}
%

{If $\dot{\gamma} = 0$, the parameter $p$ is simply the ratio between the rotational angular velocity $\dot{\delta}$ and the mean motion $n  = 2 \pi/T$. Since the average rotational velocity is inversely proportional to the time that it takes  the planet to complete a rotation around its axis, $p$ is the ratio of the orbital period over the rotational period. Therefore, {for $p = 3/2$,} the orbital period is longer than the rotation period of the planet around its axis; in particular, the planet completes a revolution around its axis in a time that corresponds to $2/3$ of its year. This is equivalent to saying that the planet completes three revolutions around its axis every two {of its} years. Figure~\ref{fig:so3to2} shows a stroboscopic view of two orbits of the planet around the star at the configuration that corresponds to the minimum of the potential for  the $p = 3/2$ spin-orbit resonance. In the view shown in the figure, the planet orbits the star and rotates around its axis in a counterclockwise direction. Since the minimum of the potential occurs at $\gamma = 0$, if the planet is at perihelion, where $\phi = M =0$, also the angle $\delta$ should be zero; this particular instant in time is labeled by \raisebox{.5pt}{\textcircled{\raisebox{-.9pt} {1}}} in the left panel of Figure~\ref{fig:so3to2}. The red line drawn on the ellipse representing the planet shows the semi-major axis that is used to measure the angle of rotation of the planet around its axis: The angle of rotation $\delta$ is the angle between the red line and a line parallel to the major axis of the orbit.  In the first year, the planet must complete one and a half revolutions around its axis, \emph{i.e.} {it} must rotate by an angle $\delta  = 3/2 \pi$ every half a year. For this reason, at aphelion the red line drawn on the planet in Figure~\ref{fig:so3to2} is perpendicular to the line (not shown in the figure)  that joins the planet to the star, as shown at the point labeled \raisebox{.5pt}{\textcircled{\raisebox{-.9pt} {4}}} in the left panel. After one year, when the planet returns to perihelion, the red line lies along the line joining the planet to the star, but it is pointing toward the star, as shown by the point labeled by \raisebox{.5pt}{\textcircled{\raisebox{-.9pt} {7}}} in the right panel of the figure, rather than away from the star as at point \raisebox{.5pt}{\textcircled{\raisebox{-.9pt} {1}}}. At the end of the second year instead, at perihelion the planet returns to the same configuration that it had at the beginning of the period shown in the left figure, with the red line parallel to the line joining the planet to the star but pointing away from the star.}

The other simple pole that appears at order $\epsilon$ in Eq.~(\ref{eq:expSandR}), $p = 1/2$ does not correspond to a minimum, but to a maximum of the function $F$ in $\gamma = 0$.
{This is due to the fact that the pole at $p = 1/2$ is the only pole among the ones explicitly written down in Eq.~(\ref{eq:expSandR}) whose coefficient at the lowest order in $\epsilon$ is positive rather than negative. Indeed, by expanding the trigonometric functions in Eq.~(\ref{eq:F})  for $p \to k/2$ with $k \in {\mathbb N}$, one finds
\begin{equation}
\sin \left(2 p \pi \right) \cos\left(2 \left( p \pi + \gamma \right) \right) = 2 \pi  \left( p - \frac{k}{2} \right)
\cos \left(2 \gamma \right) + {\mathcal O} \left(\left( p - \frac{k}{2} \right)^2 \right) \, .
\end{equation}
By multiplying this expansion by $R$ and then setting $p=k/2$, one finds that the function $F$ is proportional to \begin{displaymath}
2 \pi {\operatorname{Res}}R \left.\left(p,\epsilon \right)\right|_{p = \frac{k}{2}}  \cos \left(2 \gamma \right) \, .
\end{displaymath}
Consequently, for $p = k/2$, the function $F$ shows a minimum at $\gamma=0$ if the coefficient of the single pole at $p = k/2$ is negative, and a maximum if the coefficient is positive.
}
For the value $p =1/2$ the function $F$ has a minimum at $\gamma = \pi/2$ {as shown in the right panel of Figure~\ref{fig:minimaofFgamma0}.}

\begin{figure}
    \centering
\begin{tabular}{cc}
 \includegraphics[width=0.50\textwidth]{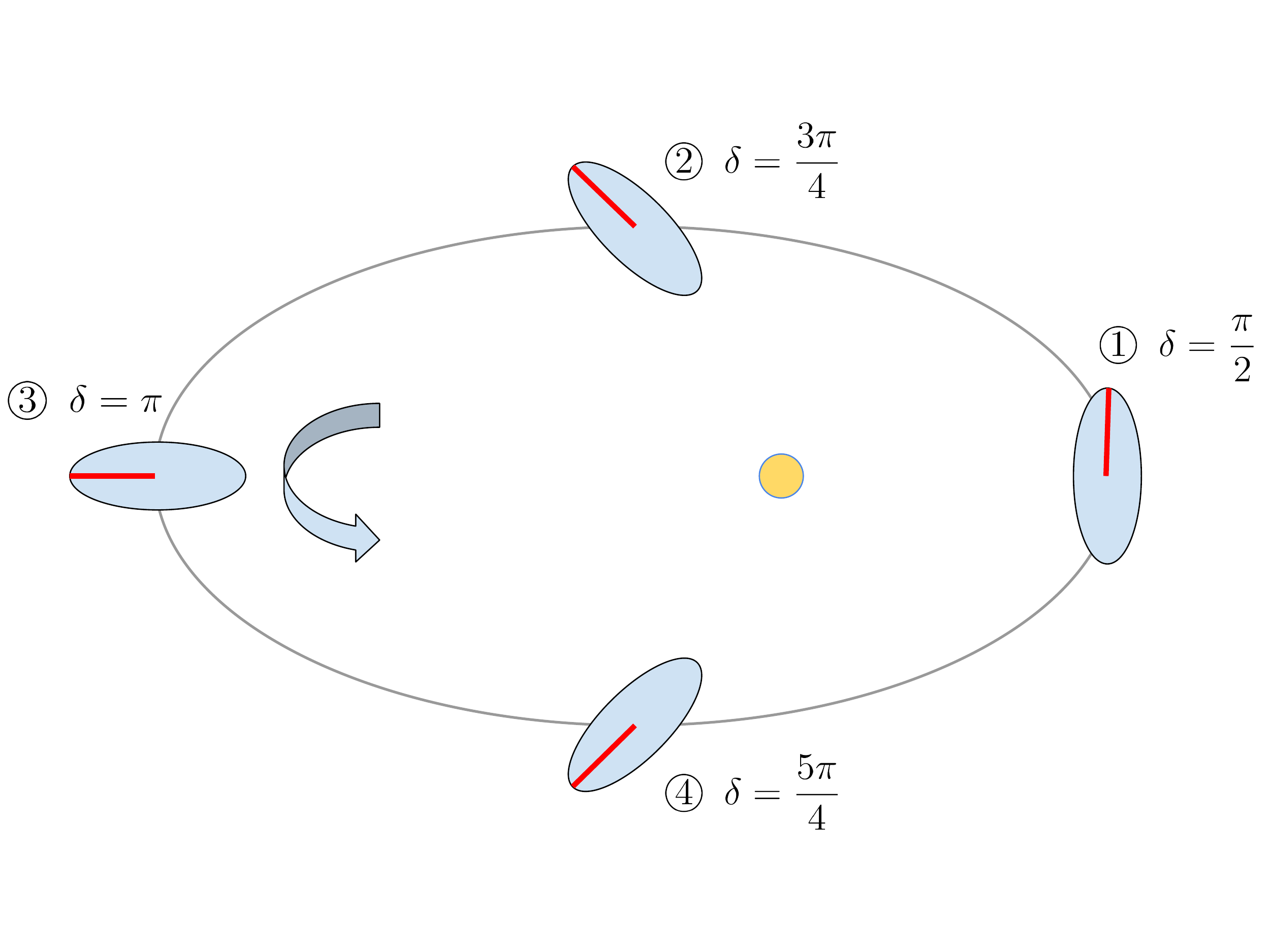}    & \includegraphics[width=0.50\textwidth]{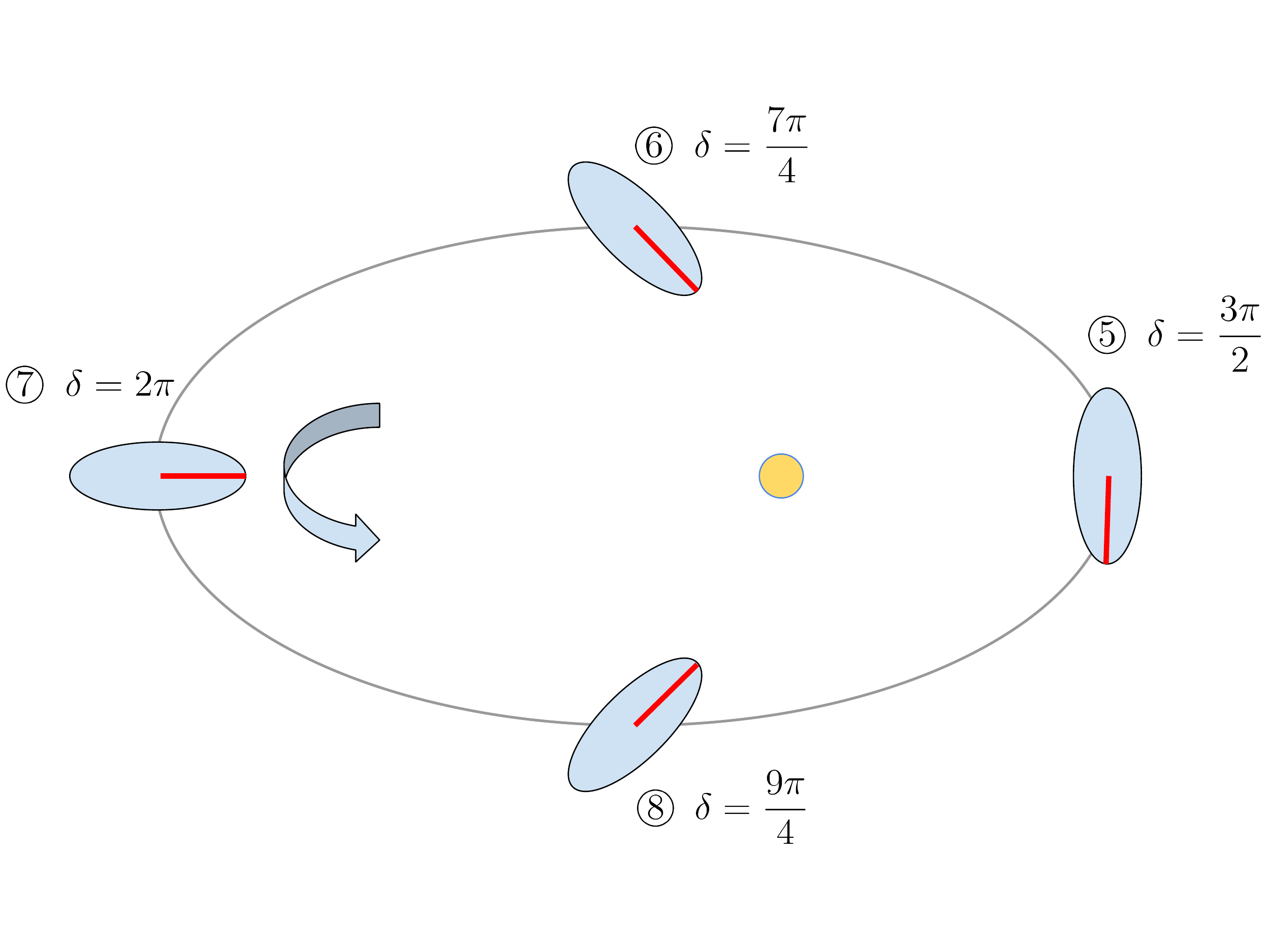}
\end{tabular}
    \caption{Schematic representation of the orbit of a planet around the star in a 1:2 spin-orbit resonance. The left panel refers to the the first orbit of the planet around the star, the right panel shows the second orbit of the planet around the star. A semi-major axis of the planet is drawn in red in order to show the angle of rotation of of the planet around its axis.}
    \label{fig:so1to2}
\end{figure}

As discussed above, this situation corresponds to the case in which the orbital mean anomaly is out of phase with respect to the rotation angle {$\delta$} by $\pi/2$, \emph{i.e.} the shortest axis of the planet $b$ points toward the Sun at perihelion. {In this configuration, the planet completes a full revolution around its axis every two years, as shown in Figure~\ref{fig:so1to2}. Since in this configuration $\gamma = \pi/2$, the value of the rotation angle $\delta$ at perihelion , \emph{i.e.} $\phi = M = 0$, should also be equal to $\pi/2$, as shown at the position labeled by \raisebox{.5pt}{\textcircled{\raisebox{-.9pt} {1}}} in the left panel of Figure~\ref{fig:so1to2}. The red line is therefore perpendicular to the line joining the star to the planet at that point. The planet rotates by an angle $\delta = \pi/2$ every half a year. For this reason the red line is parallel to the line joining the star to the planet at aphelion (points \raisebox{.5pt}{\textcircled{\raisebox{-.9pt} {3}}} and \raisebox{.5pt}{\textcircled{\raisebox{-.9pt} {7}}} in Figure~\ref{fig:so1to2}). After completing the first orbit (point \raisebox{.5pt}{\textcircled{\raisebox{-.9pt} {5}}}
in the right panel) the red line is again perpendicular to the line joining the planet to the star, but pointing in the opposite direction with respect the initial position. Also in this case, the planet returns to the initial configuration after two orbits.}

\section{Conclusions}
\label{sec:conclusions}

This work revisits the spin-orbit resonances of a planet orbiting a star in an elliptic orbit. A pedagogical approach is employed to show that for an ellipsoidal planet, the quadrupole correction to the two body gravitational potential implies that several stable configurations in which the planet rotates around its axis an integer number of times for every two revolutions around the star are possible. Among these situations, the most energetically favored is the one in which the planet spins around its axis exactly once for every revolution around the star.   This situation corresponds to the well known tidal-locking phenomenon and applies not only to a star-planet system but to any two-body gravitationally bound system. Indeed this tidal locking occurs in the Moon-Earth system and it is the reason why the Moon has a far side always hidden from Earth. A planet can be tidally locked to the star in a 1:1 spin-orbit resonance  even if the planet is a perfect sphere. Spin-orbit resonances characterized by other ratios can manifest themselves only if the orbit of the planet has a non-negligible eccentricity and the planet has a non-perfectly spherical shape.

An analysis of an appropriately defined effective potential for the rotation of the ellipsoidal  planet around its axis reveals that the second energetically most favored spin-orbit resonance is the one in which the planet spins three times around its axis for every two orbits around the star. 
In this configuration, the longest semi-axis of the ellipsoidal planet is always aligned with the major axis of the elliptic orbit at perihelion and perpendicular to it at aphelion.
Also this situation is observed in nature and indeed it describes the orbit of Mercury around the Sun, which is in fact locked in a 3:2 spin-orbit resonance. The large eccentricity of Mercury's orbit in comparison to the other planets of the solar system and  Mercury's ellipsoidal shape make this resonance more stable for Mercury than for other planets.
Similar, but energetically less favored, resonances are possible for other integer or half-integer ratios between the rotational and orbital period of the planet, such as 2:1, 5:2, etc. A spin-orbit resonance characterized by a 1:2 ratio is instead possible when the longest semi-axis of the ellipsoidal planet is perpendicular to the line joining the planet to the star at perihelion, and parallel to it at aphelion.

In this paper, the spin-orbit resonances are investigated both by means of the equation of motion for a suitably defined angle $\gamma$, as well as through the study of the shape of a potential energy term for the same angle $\gamma$. The presence of the spin-orbit resonances emerges in a straightforward way from the study of the potential. This study  could be easily incorporated in a Classical Mechanics class for physics-major undergraduate students, and it would allow the instructor to provide the students with an application of the multipole expansion of the gravitational potential. {The present study is complementary to the Exercise 2.19 found in Sussman and Wisdom's book, where, for the case of Mercury, the reader is asked to solve numerically the equations of motion satisfied by $\phi$ and $\delta$,  and to verify a posteriori that, with an appropriate choice of the initial conditions, the quantity $\delta - 3/2 \phi$ oscillates.}

A separate and more complicated question is how likely it is for a planet like Mercury to be captured in such a resonance. No attempt is made to study {or} answer this question in this work. For the Mercury-Sun system, this aspect was studied in detail in~\cite{correia1,correia2}: These studies require one to take into account the fact that Mercury is not a perfectly rigid body, an approach that goes beyond the scope of this work. 


\acknowledgements

The authors would like to thank Joel Weisberg for bringing the phenomenon of tidally locked spin-orbit resonances to their attention and Giovanni Ossola for discussions and suggestions, as well as a careful reading of the manuscript. The work of Christopher Clouse was sponsored by the CUNY Research Scholars Program (CRSP).




\begin{thebibliography}{5}


\bibitem{arons} A.~B.~Arons, ``Basic physics of the semidiurnal lunar tide,'' American Journal of Physics \textbf{47}, 934--937 (1979).
\bibitem{gron} O.~Gr{\o}n, ``A tidal force pendulum,'' American Journal of Physics \textbf{51}, 429--431 (1983).
\bibitem{white} G.~White, T.~Mondragon, D.~Slaughter, and D.~Coates, ``Modelling tidal effects,'' American Journal of Physics \textbf{61}, 367--371 (1993).
\bibitem{withers} M.~M.~Withers, ``Why do tides exist?'' The Physics Teacher \textbf{31}, 394--398 (1993).
\bibitem{koenders} M.~A.~Koenders, ``The effects of tidal forces on an elastic satellite in a closed orbit,'' European Journal of Physics \textbf{19}, 265--270 (1998).
\bibitem{butikov} E.~I.~Butikov, ``A dynamical picture of the oceanic tides,'' American Journal of Physics \textbf{70}, 1001--1011 (2002).
\bibitem{razmi} H.~Razmi, ``On the tidal force of the Moon on the Earth,'' European Journal of Physics \textbf{26}, 927--934 (2005).
\bibitem{massi} M.~Masi, ``On compressive radial tidal forces,'' American Journal of Physics \textbf{75}, 116--124 (2007).
\bibitem{urbassek} H.~M.~Urbassek, ``Precession of the Earth-Moon system,'' European Journal of Physics \textbf{30}, 1427--1433 (2009).
\bibitem{pujol} O.~Pujol, C.~Lagoute and J.~P.~P\'erez, ``Weight, gravitation, inertia, and tides,'' European Journal of Physics \textbf{36}, 065012, 1--12 (2015).
\bibitem{ng} C.~Ng, ``How tidal forces cause ocean tides in the equilibrium theory,'' Physics Education \textbf{50}, 159--164 (2015).
\bibitem{cregg} P.~J.~Cregg, ``Just how much do the planets affect the tides?'' Physics Education \textbf{52}, 053003, 1--6 (2017).
\bibitem{norsen} T.~Norsen, M.~Dreese, and C.~West, ``The gravitational self-interaction of the Earth's tidal bulge,'' American Journal of Physics \textbf{85}, 663--669 (2017).
\bibitem{kepler} {Johannes Kepler, \emph{Astronomia nova} (1609).}
\bibitem{newton} Sir Isaac Newton, \emph{Philosophi\ae~Naturalis Principia Mathematica} (1687).
\bibitem{kant} Immanuel Kant, ``Whether the Earth Has Undergone an Alteration of Its Axial Rotation,'' W\"ochentliche Frag- und Anzeigungs-Nachricten, K\"onigsberg, (1754).
\bibitem{kopal72} Z. Kopal, ``Tidal Evolution in Close Binary Systems,'' Astrophysics and Space Science \textbf{17}, 161--185 (1972).
\bibitem{counselman73} C. C. Counselman, ``Outcomes of Tidal Evolution,'' Astrophysical Journal \textbf{180}, 307--314 (1973).
\bibitem{vanhamme79} W. van Hamme, ``On Synchronism Between Axial Rotation and Orbital Motion in Close Binary Systems,'' Astrophysics and Space Science \textbf{64}, 239--248 (1979).
\bibitem{hut80} P. Hut, ``Stability of Tidal Equilibrium,'' Astronomy and Astrophysics \textbf{92}, 167-170, (1980). 
\bibitem{mcdonald} http://www.hep.princeton.edu/$\sim$mcdonald/examples/spin\_orbit.pdf
\bibitem{ferroglia} A.~Ferroglia and M.~C.~N.~Fiolhais, ``Tidal locking and the gravitational fold catastrophe,'' American Journal of Physics \textbf{88}, 1059--1067 (2020).
\bibitem{guemez} J.~G\"{u}\'emez, C.~Fiolhais, and M.~Fiolhais, ``The Cartesian diver and the fold catastrophe,'' American Journal of Physics \textbf{70}, 710--714 (2002).
\bibitem{fiolhais1} M.~Fiolhais and R.~Nogueira, ``Sistema mec\'anico con un potencial catastr\'ofico,'' Revista Espa\~nola de F\'isica Vol \textbf{34}, No 1, 30--33 (2020).
\bibitem{fiolhais2} M.~Fiolhais, B.~Golli and R.~Nogueira, ``Mechanical apparatus for the fold catastrophe demonstration,'' European Journal of Physics \textbf{42}, 045001 (2021).   
\bibitem{dyce} R.~B.~Dyce, G.~H.~Pettengill, I.~I.~Shapiro, ``Radar determination of the rotations of Venus and Mercury,'' 
The Astronomical Journal \textbf{72}, 351–359 (1967).
{\bibitem{sussman} G.~J.~Sussman, J.~Wisdom, ``Structure and Interpretation of Classical Mechanics,'' The MIT press (2015).}
\bibitem{taylor} J.~R.~Taylor, \emph{Classical Mechanics} (University Science Books, Mill Valley, 2005). See section~8.6.
{\bibitem{barger} V.~Barger and M.~Olsson, \emph{Classical Mechanics: A Modern Perspective} (McGraw-Hill, Inc., New York, 1995). See equation 5.64 on page 148.}
\bibitem{goldreich_peale} P.~Goldreich and S.~Peale, ``Spin-Orbit Coupling in the Solar System,'' Astronomical Journal \textbf{71},  425-438, (1966).
\bibitem{Montenbruck} O.~Montenbruck, \emph{Practical Ephemeris Calculations} (Springer-Verlag, New York, 1989). See page 44.
\bibitem{Meeus} J.~Meeus, \emph{Astronomical Algorithms} (Willmann-Bell Inc., Richmond, VA, 1991). See page 182.
\bibitem{danby} J.~M.~A. Danby,  \emph{Fundamentals of Celestial Mechanics} (Willmann-Bell Inc., Richmond, VA, 1962).
\bibitem{correia1} A.~C.~M.~Correia and J.~Laskar, ``Mercury's capture into the 3/2 spin-orbit resonance including the effect of core-mantle friction,'' Icarus \textbf{201}, 1--11 (2009).
\bibitem{correia2} A.~C.~M.~Correia and J.~Laskar, ``Long-term evolution of the spin of Mercury. I. Effect of the obliquity and core-mantle friction,'' Icarus \textbf{205}, 338--355 (2010).



\end{thebibliography}
\end{document}